\documentclass[fleqn,12pt,twoside]{article}
\usepackage[headings]{espcrc1}

\readRCS
$Id: espcrc1.tex,v 1.2 2004/02/24 11:22:11 spepping Exp $
\usepackage{graphicx}
\usepackage[figuresright]{rotating}

\newcommand{\AmS}{{\protect\the\textfont2
  A\kern-.1667em\lower.5ex\hbox{M}\kern-.125emS}}

\title{Dissipation and memory effects in pure glue deconfinement}

\author{E. S. Fraga\address[IF-UFRJ]{Instituto de F\'\i sica, 
Universidade Federal do Rio de Janeiro, \\ Caixa Postal 68528, 
Rio de Janeiro, RJ 21941-972, Brazil}, T. Kodama\addressmark[IF-UFRJ], 
G. Krein\address[IFT]{Instituto de F\'\i sica Te\'orica, 
Universidade Estadual Paulista, \\ Rua Pamplona 145, 
S\~ao Paulo, SP 01405-900, Brazil},
A. J. Mizher\addressmark[IF-UFRJ], and L. F. Palhares\addressmark[IF-UFRJ]}
       
\runtitle{Dissipation and memory effects in pure glue deconfinement}
\runauthor{E.S. Fraga, G. Krein, T. Kodama, A.J. Mizher and L.F. Palhares}

\begin{document}

\maketitle

\begin{abstract}
We investigate the effects of dissipation in the deconfining 
transition for a pure $SU(2)$ gauge theory. Using an effective 
model for the order parameter, we study its Langevin evolution 
numerically, and compare results from local additive noise 
dynamics to those obtained considering an exponential non-local 
kernel for early times. 
\end{abstract}

\vspace{1cm}

Nowadays Lattice QCD provides a very reliable theoretical testing 
ground for the study of the phase structure of strongly interacting 
matter \cite{laermann}. Even results obtained within the simpler 
framework of pure gauge $SU(N)$ can bring some insight to the 
analysis of experimental data from high-energy heavy ion collisions 
\cite{bnl}. Inspired by lattice results, one can build effective 
field theory models to be used in the study of the dynamics of phase 
conversion in the deconfining transition of non-abelian gauge theories.

In particular, for a pure gauge $SU(N)$ theory, the Polyakov loop provides 
well-defined order parameters \cite{polyakov,thooft,pisarski,ogilvie}, 
and one can construct an effective Landau-Ginzburg field theory based on 
these quantities. The effective potential for 
$T<<T_d$ has only one minimum, at zero, where the whole system is localized. 
With the increase of the temperature new minima appear: $N$ minima 
for $Z(N)$, the center of $SU(N)$. At the critical temperature, 
$T_d$, all the minima are degenerate, and above $T_d$ the new  
minima become the true vacuum states of the theory, so that the system 
starts to decay. 

In this paper we consider the case of pure $SU(2)$. The transition being 
second-order \cite{yaffe}, there is never a barrier to overcome. The process 
of phase transition will proceed through spinodal decomposition, which will 
have its exponential explosion retarded by the effect of dissipation in 
the medium, as was shown for the case of the chiral transition in 
Ref. \cite{Fraga:2004hp}. Using the effective model proposed in Ref. 
\cite{ogilvie}, we study the Langevin evolution of the order parameter, 
comparing results, for early times, from local additive 
noise dynamics to those obtained considering non-markovian effects 
brought about by an exponential non-local kernel. The choice of the kernel 
is such that one can trade off the memory integral for higher-order time 
derivatives in the differential equation for the order parameter. The 
simple form of the kernel does not restrict our semi-quantitative 
analysis since the only physically significant parameter in the 
memory kernel is its width, so that different choices for its functional 
form should not modify the output appreciably.

The effective theory we adopt \cite{ogilvie} is based on a mean 
field treatment in which the Polyakov loops are constant throughout the 
space and the free energy is a function of its eigenvalues, 
$P_{jk} = \exp(i\theta_j) \; \delta_{jk}$. To a purely perturbative 
computation of the free energy of gluons in $SU(N)$ 
one adds, phenomenologically, a mass scale $M=M(T_d)$ to the 
dispersion relation, {i.e.} $\omega_k=\sqrt{k^2+M^2}$. $T_d$ 
is the critical temperature for deconfinement that can be read off 
from lattice data. For $SU(2)$, $T_d=302~$MeV \cite{karsch}.
Parametrizing the Polyakov loop as $diag[\exp(i\phi,-i\phi)]$ 
and defining a more convenient variable $\psi\equiv 1-\pi\phi/2$, 
which plays the role of the order parameter for deconfinement, 
one obtains:
\begin{equation}
V_{eff} = -\frac{\pi^2T^3}{15}+\frac{T^3\pi^2}{12}(1-\psi^2)^2 
+ \frac{M^2T}{4} -\frac{M^2T}{4}(1-\psi^2) \; ,
\end{equation}
so that $\langle\psi\rangle=0$ in the confined phase and 
$\langle\psi\rangle\to \psi_0$ in the 
deconfined phase. One can easily connect the behavior of 
$\psi$ to that of the trace of the Polyakov loop, 
also used as an order parameter \cite{pisarski}, via 
${\rm Tr} \, L = 2 \cos[\pi (1-\psi)/2]$. The value of $M$ 
can be determined from the deconfinement temperature through 
the relation $T_d=(3/2)^{1/2}M/\pi \approx 0.38985M$, and  
the minima of the potential occur at 
$\psi_0=\pm \left( 1 - T_d^2/T^2 \right)^{1/2}$.
\begin{figure}[htb]
\begin{minipage}[t]{75mm}
\includegraphics[width=\linewidth]{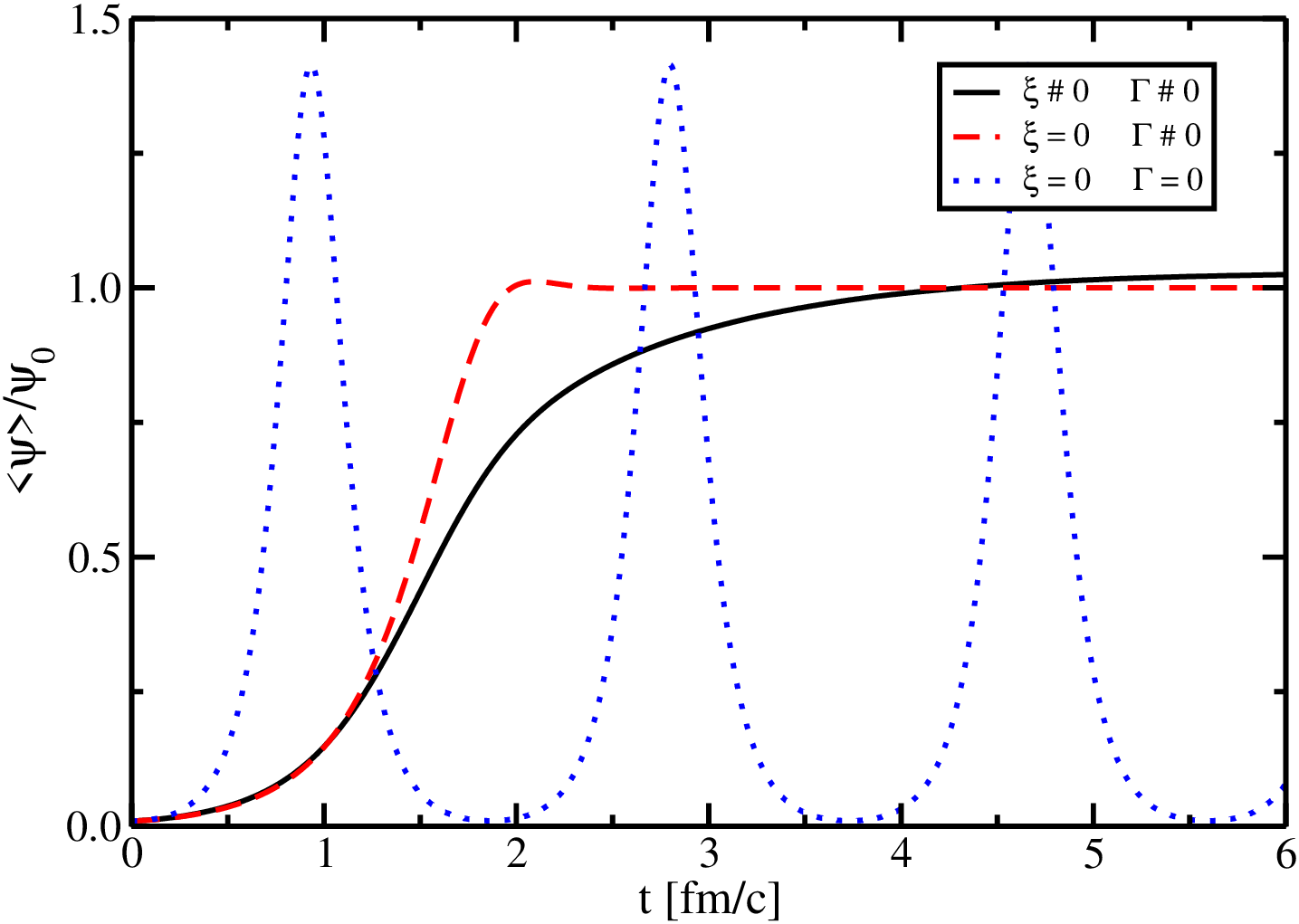}
\caption{Volume average of the $SU(2)$ order parameter normalized 
by the positive minimum of the bare effective potential.}
\end{minipage}
\hspace{\fill}
\begin{minipage}[t]{75mm}
\includegraphics[width=\linewidth]{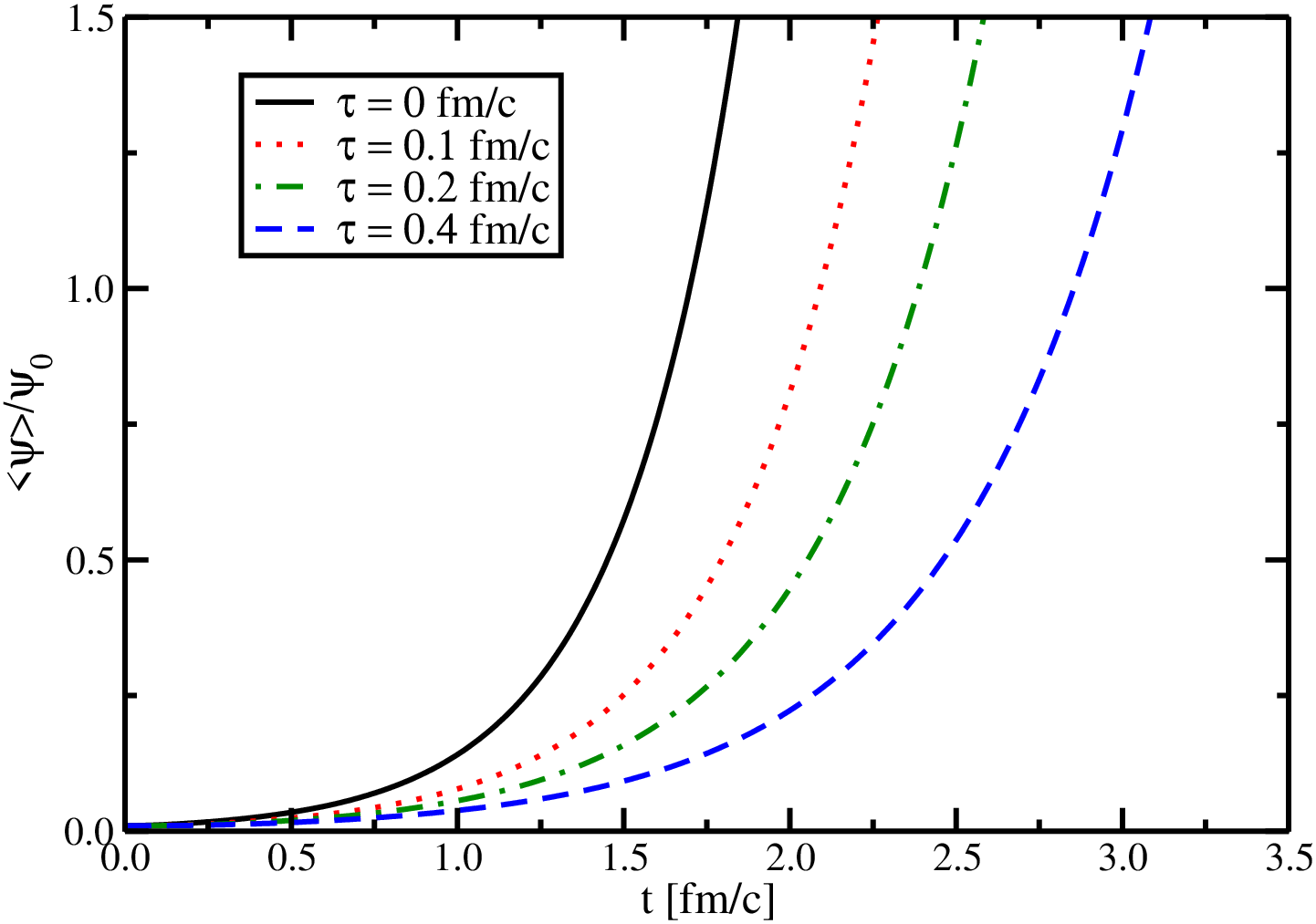}
\caption{Comparison of the early time (exact) evolution of the 
$SU(2)$ order parameter for different values of $\tau$.}
\end{minipage}
\label{fig1}
\end{figure}

Let us now consider the nonequilibrium evolution of the order 
parameter for the spontaneous breakdown of $Z(2)$. As costumary, 
we assume the system to be characterized by a coarse-grained free 
energy in the Landau-Ginzburg fashion
\begin{equation}
F_{cg}([\psi],T)=\int d^3 x\left[\frac{\sigma(T)}{2} \, 
(\nabla \psi)^2+
V_{eff}([\psi],T)\right] \; ,
\end{equation}
where $V_{eff}([\psi],T)$ is the effective potential obtained 
previously, $\sigma(T) = \pi^2 T/g^2$ plays the role of a surface 
tension and $g=g(T)$ is the gauge coupling.

A markovian description of the time evolution of the order parameter 
and its approach to equilibrium can be implemented through the 
traditional additive-noise Langevin equation
\begin{equation}
\sigma(T)\left[\frac{\partial^2\psi}{\partial t^2} 
- \nabla^2\psi \right]  
+ \Gamma(T) \, \frac{\partial\psi}{\partial t} + V'_{eff}([\psi],T) = 
\xi (\vec x, t) \; .
\label{langevin}
\end{equation}
The function $\xi$ is a stochastic noise assumed gaussian and white 
so that $\langle\xi (\vec x, t)\rangle=0$ and 
$\langle \xi (\vec x, t)\xi(\vec x' ,t')\rangle=
2\Gamma \delta (\vec x- \vec x' )\delta (t - t')$. 
$\Gamma$ is a kinetic coefficient that we fix in the following way. 
Performing pure-gauge euclidean lattice Monte Carlo simulations in the 
line discussed in Ref.~\cite{ogilvie}, spinodal decomposition 
is obtained through local heat-bath updates of gauge field configurations 
at $\beta = 4/g^2 =3$ (corresponding to $T = 6.6 \, T_d$), 
after thermalizing the lattice at $\beta = 4/g^2 = 2$. 
The critical value for deconfinement is found to be $\beta_d \sim 2.3$. 
$\Gamma$ is then extracted by comparing the short-time exponential growth of 
the two-point correlation function predicted by the 
simulations \cite{Krein:2005wh} to the Langevin description assuming, 
of course, that both dynamics are the same (see, also, the extensive 
studies of Berg {\it et al.} \cite{berg}). Assuming that typical 
thermalization times are of the order of a 
few fm/c, one can relate Monte Carlo time and real time and obtain
$\Gamma \sim 10^3$~fm$^{-2}$. 

To incorporate memory effects in our description, we consider the 
generalized time-dependent Landau-Ginzburg evolution equation 
\cite{bray,Koide:2006vf}
\begin{equation}
\sigma(T)\frac{\partial^2\psi}{\partial t^2} +
\Gamma(T) \, \frac{\partial\psi}{\partial t}=
\int_0^t dt'~\Sigma(t-t')
\left[ -\frac{\delta F([\psi],T)}{\delta\psi(t')} + 
\xi(t') \right] \, .
\end{equation}
For analytical purposes, it is convenient to choose for 
the kernel $\Sigma(t-t')=e^{-(t-t')/\tau}/\tau$, which is a 
representation of the Dirac delta function in the limit 
$\tau\to 0$, in which the description becomes markovian. 
Different functional forms for a localized kernel that 
yields the markovian result in the appropriate limit are 
essentially equivalent, the only relevant scale being the 
width $\tau^{-1}$. For the particular choice above, one 
can dispose of the memory integral in favor of an additional 
third-order derivative in time, obtaining the following evolution 
equation for the order parameter:
\begin{equation}
\tau\sigma\frac{\partial^3\psi}{\partial t^3} + 
\left[ \sigma + \tau\Gamma(T) \right] \frac{\partial^2\psi}{\partial t^2} +
\Gamma\frac{\partial\psi}{\partial t} - 
\sigma(T)\nabla^2\psi + V'_{eff}([\psi],T)
= \xi(\vec{x},t) \, ,
\label{memory}
\end{equation}
which reproduces (\ref{langevin}) in the limit of vanishing $\tau$.

We solve Eq. (\ref{langevin}) numerically on a lattice, 
average over several realizations with random initial configurations 
around $\psi \sim 0$, and study the evolution of the volume average 
$\langle \psi \rangle$ normalized by $\psi_0$.
In Fig. 1 we plot $\langle \psi \rangle$, normalized by the 
positive minimum of the bare effective potential, 
for three situations: no dissipation and no noise (dotted curve), 
markovian dissipative evolution with no noise (dashed curve) and 
with white noise (solid curve). When considering noise, we have 
added the appropriate counterterms to make the equilibrium 
solution independent of the lattice spacing~\cite{CT}. 
From this comparison, it is clear that dissipation and noise 
delay appreciably the thermalization time scale for the deconfining 
transition as was discussed in \cite{Mizher:2006zg}.

In Fig. 2, we show results for the early-time (exact) solution of 
Eq. (\ref{memory}) for very conservative values of the width 
$\tau^{-1}$. Here, we have linearized the effective equation of 
motion and solved it for the homogeneous average of 
$\psi$ at early times, a process which allows for treating noise 
exactly  \cite{Palhares:2006wt}. From Fig. 2, it 
is clear that even for small values of the reaction time the delay 
for the onset of the deconfining transition is quite significant. 
Although our analysis is restricted to the short-time behavior 
(up to $\sim 1.5~$fm/c) of 
the order parameter in a pure gauge theory, it indicates that 
one should include non-markovian effects in more realistic 
approaches to the phenomenology of high-energy heavy ion 
collisions. For instance, large modifications brought about by 
memory kernels in a nonequilibrium calculation of dilepton production 
seem to play an important role in understanding the 
data \cite{Schenke:2006uh}. Strictly speaking, Eq. (\ref{memory})
does not have a parabolic structure and may exhibit causality 
problems similar to the ones found in the usual diffusion 
equation \cite{Koide:2005qb}. 
On the other hand, the inclusion of retardation effects as discussed 
above was shown to be crucial for satisfying causality constraints in 
relativistic dissipative hydrodynamics, and may provide an alternative 
to a second-order formulation of thermodynamics \cite{Kodama:2006ga}.

We thank G. Ananos, A.~Bazavov, A. Dumitru, T. Koide 
and P. Romatschke for discussions. 
This work was partially supported by 
CAPES, CNPq, FAPERJ, FAPESP and FUJB/UFRJ. 


\end{document}